\documentstyle[preprint,revtex,eqsecnum]{aps}

\begin{document}
\draft
\preprint{Alberta-Thy-40-92}
\begin{title}
Cauchy horizon singularity without mass inflation
\end{title}
\author{Patrick R. Brady$^{\dagger}$, Dar\'\i o
N\'u\~nez$^{\dagger\ast}$ and
Sukanya Sinha$^{\ast}$}

\begin{instit}
$^{\dagger}$ Theoretical Physics Institute, University of Alberta,
Edmonton,\\
Alberta T6G 2J1, Canada. \\
$^{\ast}$ Instituto de Ciencias Nucleares, UNAM,\\ Circuito Exterior
CU, A.P.
70-543, M\'exico, D. F. 04510,  M\'exico
\end{instit}

\begin{abstract}
A perturbed Reissner-Nordstr\"om-de~Sitter solution is used to
emphasize the
nature of the singularity along the Cauchy horizon of a charged
spherically
symmetric black hole.  For these solutions, conditions may prevail
under which
the mass function is bounded and yet the curvature scalar
$R_{\alpha\beta\gamma\delta}
R^{\alpha\beta\gamma\delta}$ diverges.
\end{abstract}
\vfill
\eject
\section{Introduction}
The external gravitational field of a black hole, formed by collapse,
settles
down to the Kerr-Newman solution of General Relativity.  These
generic stationary black hole solutions have a rather curious feature, the
inner horizon
acts as the
boundary of predictability for external initial value problems.
However,
Penrose~\cite{Pen:68} pointed out that this inner (Cauchy) horizon is
a surface
of infinite blueshift, and thus partial absorption of the
radiative tail of
the gravitational collapse leads to a divergent flux of energy  along
it. This
suggests that the backreaction of perturbations on the geometry may
generate
unbounded curvature along the Cauchy horizon~(CH). Detailed
calculations, which
analysed the evolution of gravitational perturbations confirmed the
divergence
of the energy flux~\cite{S&M}, and for charged spherical black
holes Poisson
and Israel(PI)~\cite{P&I:90} showed that a scalar curvature
singularity does
form at the CH when the
above influx is accompanied by the outflux emitted from the
collapsing star.\  Since they found the mass function diverges, they
called it a
 {\em mass inflation} singularity. An exact solution, based on the
work of PI
was used by Ori to examine the nature of the singularity in some
detail~\cite{Ori:91}, and he showed that an observer, who falls
into the black hole, experiences finite tidal distortion at this
singularity.

The presence of a cosmological constant (of arbitrary magnitude)
modifies this
picture in a non-trivial manner, and serves to emphasize an aspect of
the
singularity along the CH which has so far been ignored.  The leading
divergence
of  the curvature scalars is not contained in the Weyl scalar, but rather in
$R_{\alpha\beta\gamma\delta}
R^{\alpha\beta\gamma\delta}$.

 For a generic perturbation of a Reissner-Nordstr\"om black hole in
de~Sitter
space, the CH can be stable for some values of the charge~($e$) and the
mass~($m$)~\cite{B&P:92,M&M}.  These configurations are quite rare
though, and require $|e| > m$.  In~\cite{B&P:92} it was speculated that this
spacetime might provide a counter-example to the conjecture that non-scalar
curvature
singularities are catalyzed into stronger scalar curvature
singularities, once
perturbed.  This was based on the observation that the Weyl curvature
of the
solution to the Einstein field equations (for simple spherical
perturbations)
is dominated  by the proper time integral of the energy flux along the
horizon.
 It was shown in~\cite{B&P:92} that this flux behaves like ${\cal F}(v) \sim
e^{2(\kappa_{\rm i} - \kappa_{\rm c})v}$ and
   \begin{equation}
\Psi_2 \propto   \int {\cal F}(v) e^{\kappa_{\rm i}v}dv \sim
e^{(\kappa_{\rm
i} -
   2\kappa_{\rm c})v}\: ,
   \end{equation}
where $v\rightarrow \infty$ on the CH. The constants $\kappa_{\rm i}$ and
$\kappa_{\rm c}$ are the surface gravities of the inner and cosmological
horizons respectively.
Thus one finds values of the charge and the external mass of the
black hole for
which $\kappa_{\rm c} < \kappa_{\rm i} < 2\kappa_{\rm c}$, and the
observed
energy density grows without bound, but $\Psi_2$ is finite.

Although these observations are correct, a scalar curvature
singularity does
form along the CH contrary to the speculation in~\cite{B&P:92}.  We will
demonstrate this fact using a model in
which
perturbations of a Reissner-Nordstr\"om-de~Sitter black hole are
modelled by
crossflowing streams of null dust, as in~\cite{P&I:90}.
The singularity is characterized by the divergence of
$R_{\alpha\beta} R^{\alpha\beta}$, where the
dominant
behavior is proportional to the influx of blueshifted radiation.
We wish to
emphasize that this behavior should be present  in all spherically
symmetric
solutions, where the outflux from the star is continuous, and does
not require
a non-zero cosmological constant.  However, the  presence of a
cosmological
term shows clearly that the divergence of the mass function is not
{\em
necessary} for the presence of a scalar curvature singularity.

\section{An approximate solution}

The spherically symmetric line element can be written as
\begin{equation}
ds^2= d\sigma^2+r^2d\Omega^2,\mbox{\hspace{.25in}}d\sigma^2 =
g_{ab}dx^adx^b,
\end{equation}
where $d\Omega^2$ is the
line element on the unit two-sphere, and latin indices $a,b,\ldots$
range over
(0,1).  Along with the scalar function $r(x^a)$ we introduce
$m(x^a) , f(x^a)$ and $\kappa(x^a)$ defined by
\begin{eqnarray}
g^{ab}r_{,a}r_{,b} & = & f(x^a) = 1 -\frac{2 m(x^a)}r +
\frac{e^2}{r^2}
-\frac{\Lambda}3 r^2, \label{eqn:3} \\
\kappa & = & -\frac{1}2 \partial_r f.
\end{eqnarray}
It is also convenient to use null coordinates $U$, $V$ on the
``radial'' two
spaces so that $d\sigma^2 = -2e^{2\nu}dUdV$ and $\nu = \nu (U,V)$.
It is now
easy to generalize the derivation of the field equations
in~\cite{P&I:90} to
include a cosmological constant.  The result is
\begin{eqnarray}
r_{;ab} + \kappa g_{ab} &=&  - 4\pi r T_{ab}\: , \label{eqn:5}\\
\Box \nu &=& -\partial_r \kappa  = -{1\over r^3}\left(2m - {3e^2\over
r}
+{\Lambda\over 3} r^3 \right)\: , \label{eqn:6}\\
\Box m &=& - (4\pi)^2 r^3 T_{ab}T^{ab} \label{eqn:7}\: ,
\end{eqnarray}
where a semi-colon indicates the covariant derivative associated with
$d\sigma^2$, and $T_{ab}$ is the 2$\times$2 sub-matrix  of the
stress-energy
tensor.

For crossflowing null dust, the stress-energy tensor is
\begin{equation}
T_{\mu\nu} = {\rho}_{\rm in} l_{\mu} l_{\nu} + {\rho}_{\rm out}
n_{\mu} n_{\nu}
\: , \label{eqn:8}
\end{equation}
where $l_{\mu} = -\partial_{\mu} V$ and $ n_{\mu} =  -\partial_{\mu}
U$ are
radial null vectors pointing inwards and outwards respectively, and,
$\rho_{in}$ and $\rho_{out}$ represent the energy
densities of the inward and outward fluxes.  Covariant conservation
requires
that they are given by
\begin{equation}
\rho_{\rm in} = \frac{L_{\rm in}(V)}{4\pi r^2} \: ,
\mbox{\hspace{.25in}}
\rho_{\rm out} = \frac{L_{\rm out}(U)}{4\pi r^2} \: .
\end{equation}

Let us imagine that the inflow is turned on at a finite
advanced time $V= V_0$ and the outflow is turned on at $U = U_0$.
In the pure inflow(outflow) regime
the solution is an ingoing(outgoing)
Vaidya-Reissner-Nordstr\"om-de~Sitter spacetime with mass function
$m(V)(m(U))$.
The global structure of the spacetime with cross-flow
is shown in the Penrose diagram of Fig. 1.
The loci of the three apparent horizons, in the pure inflow region,
are given
by the positive roots of $f=0$.  The ingoing sheet of the inner
horizon is
$r=r_{\rm i}$, and of the cosmological horizon is $r=r_{\rm c}$.  The outer
apparent
horizon  settles down to the constant radius $r=r_{\rm e}$, for
asymptotically large
$v$, and the function $\kappa$ evaluated at a particular horizon
gives the
surface gravity  of that horizon.

We choose $V$ such that $V=0$ at the Cauchy horizon. Note that this
does not
specify the coordinate completely,
but ensures  that it may be  regularly related to the Kruskalized
advanced time associated with the inner apparent horizon.
Our eventual aim is to construct an approximate solution which is
valid as
$V\rightarrow 0$.

Taking the trace of (\ref{eqn:5}) we find that $r(U,V)$ also
satisfies
\begin{equation}
\Box r = -2 \kappa \label{eqn:10}\: .
\end{equation}
Using (\ref{eqn:3}), (\ref{eqn:6}) and (\ref{eqn:10}) we obtain the
wave
equation
\begin{equation}
\Box (ln (re^{2\nu}))  = {1\over r^4}(3 e^2 - r^2 - \Lambda
r^4)\label{eqn:11}
\: ,
\end{equation}
which can be integrated formally to
\begin{equation}
re^{2\nu} = \alpha g_1(U) g_2(V)\exp{\left[ F(U,V) \right]}\: ,
\label{eqn:12}
\end{equation}
with
\begin{equation}
 F(U,V) = \frac{1}{2} \int_{U_0}^{U}\!\!\int_{V_0}^{V}\!\!dU'dV'
 \frac{e^{2\nu'}}{(r')^4} \left(
(r')^2-3 e^2  +
\Lambda (r')^4\right)  \: . \label{eqn:13}
 \end{equation}
The functions $g_1(U)$ and $g_2(V)$ are determined by the boundary
conditions
along the null rays $V=V_0$ and $U=U_0$, and $\alpha$ is a constant
with the
dimension of length.  In~\cite{Waugh:} and~\cite{P&I:90} it was shown
that null
coordinates exist such that $g_2(V)$ is well behaved as $V\rightarrow
0$,  in
particular we can rescale the coordinates so that $g_2(0) = 1$.  This
guarantees that $V= - \mbox{\rm (const.)} \times e^{- \kappa_{\rm i} v}$,
where
$ \kappa_{\rm i}$ is the surface gravity of the static portion of the inner
horizon.
Similarly, provided we limit ourselves to values of $U$ near to $U_0$
we may
set $g_1(U)=1$ without loss of generality.

In order to proceed,  we must estimate the behavior of the integral
(\ref{eqn:13}).  The crucial observation (which was made
in~\cite{P&I:90} but
not carried to its full consequences) is that the integrand is
negative for
$r\le r_{\rm i}$, and we expect the leading behavior to come from near
the CH.
Thus if it diverges, the integral must diverge to negative values and
$re^{2\nu} \rightarrow 0$.  As an immediate consequence of this, we
see from
(\ref{eqn:13}) that this would require that $r$ go to zero. Since  we
expect
$r(U,V)$ to be a well behaved function, with the {\em slow}
contraction of
ingoing null rays governed primarily by the outflux from the star,
we conclude
 that a good approximation to the metric coefficient is
\begin{equation}
e^{2\nu} \simeq \frac{\alpha\exp\left[ F(U,0) \right]}{r}\:
,\label{eqn:14}
\end{equation}
near to $V=0$.

With this ansatz we can rewrite (\ref{eqn:5}) as
\begin{eqnarray}
(r^2)_{,UU} - F_{,U} (r^2)_{,U} &\simeq& -2L_{\rm out}(U) \\
(r^2)_{,VV}  &\simeq& -2L_{\rm in}(V) \: ,
\end{eqnarray}
where $F$ is treated as a function of $U$ alone from here on.
The solution of these equations is then given by
\begin{equation}
r^2 \simeq r_{i}^2 -
2\int_{V_0}^{V}\!\!dV'\left[\int_{V_0}^{V'}\!\!dV''
L_{\rm in}(V'')\right]
 + \int_{U_0}^{U}\!\! \!e^{F'}\left[ V -
2\int_{U_0}^{U'}\!\!\!e^{-F''}L_{\rm out}(U'')dU''
\right] dU'\: .\label{eqn:15}
\end{equation}
We still have not determined in detail what the function $F(U,0)$ is,
but it is
unimportant in the ensuing analysis.

Before proceeding to analyse the curvature singularity which forms
due to the
infinitely blueshifted influx, we present the solution of equation
(\ref{eqn:7}) for the mass function.  In these coordinates and using
(\ref{eqn:14}) it is easy to integrate this to
\begin{eqnarray}
m(U,V) &\simeq& \alpha^{-1} \int_{U_0}^{U}\!\!\left[ e^{-F'}L_{\rm out}(U')
\right]dU'
 \int_{V_0}^{V}\!\!\left[ L_{\rm out}(V') \right]dV' \nonumber \\ &&\ \ \ +
m(U_0,V_0)
-{\beta}
 \int_{V_0}^{V}\!\!\left[ V'L_{\rm out}(V') \right]dV'  \ \ +
{\gamma}\int_{U_0}^{U}\!\!\left[ U'L_{\rm out}(U') \right]dU'
\label{eqn:16}
\: ,
\end{eqnarray}
where $\beta$ and $\gamma$ are constants.
The contribution from the pure inflow (outflow) region is  finite for
realistic
perturbations, and it is only when both fluxes are present that the
first term
can lead to a divergence of the mass function.

The functional form of the luminosities $L_{\rm out}(U)$ and $L_{\rm in}(V)$
is
as yet
unspecified, so we now specialize to the case of interest.
In~\cite{B&P:92}
and~\cite{M&M} it was shown that for a generic perturbation the
initial
conditions on the influx should be
\begin{equation}
L(V) \sim K(V)[- V / \alpha]^{\frac{2(\kappa_{\rm c}-\kappa_{\rm
i})}{\kappa_{\rm i}}}
\: ,
\label{eqn:16+}
\end{equation}
near to the CH.  This is proportional to the energy density of the
influx, as
measured by a free falling observer crossing the CH in the pure
inflow region.
Notice that it diverges provided $\kappa_{\rm i} > \kappa_{\rm c}$.  The
behavior
of the
prefactor in (\ref{eqn:16+}) is unimportant, so we take
$K(V)\rightarrow {\rm
constant}$ as $V\rightarrow 0$ (if the cosmological constant is zero,
then
$ \kappa_{\rm c} = 0$ and $K(V)$ is given by the analysis of
Price~\cite{Price} as
$K(V) \sim |\ln(-V/\alpha)|^{-p}$).  Since it is slowly varying
 compared  to the other factor in (\ref{eqn:16+}), we can write
\begin{equation}
\int dV L(V)  \sim K(V){ [-V/ \alpha]}^{ {(2\kappa_{\rm c} -
\kappa_{\rm i})}\over
\kappa_{\rm i}}.
\label{eqn:17}
\end{equation}
The mass function is proportional to (\ref{eqn:17}), so it will
diverge as
$V \rightarrow 0 $ unless $2\kappa_{\rm c} > \kappa_{\rm i}$.

\section{The Cauchy horizon singularity}

Our solution, given by (\ref{eqn:14}), (\ref{eqn:15}) and
(\ref{eqn:16}), gives
conditions under which the mass function is finite,   and shows that
the metric
is well behaved provided $r$ is bounded away from zero. In this
section we
examine the nature of the singularity that is present for a general
outflux.
There are three essentially different possibilities; (i)~complete
stability
when $ \kappa_{\rm c} > \kappa_{\rm i} $, (ii)~divergent influx but finite
mass
function
when $ 2\kappa_{\rm c} > \kappa_{\rm i}> \kappa_{\rm c}$ and (iii)~both
divergent
influx and
mass function when $ \kappa_{\rm c} < \kappa_{\rm i} $.  Defining
a=$e^2/m_1^2$
and  $b=
\Lambda m_1^2/3$, where $m_1$ is the asymptotic value of the mass
function in
the pure inflow region, we have plotted  these conditions in Fig.~2.
Some
useful relations involving the surface gravities are also given in
appendix A.

We wish to focus our analysis on case~(ii) where no mass inflation
has occurred
despite the presence of the infinite energy density.   At first
sight, one is
tempted to conclude that the spacetime is regular, since we know that
the mass
function characterizes the Weyl curvature scalar $\Psi_2 \propto
m/r^3$ within
spherical symmetry.  However with a moment's thought one realizes that
Eqn.~(\ref{eqn:8}) implies that $R_{\alpha\beta}
R^{\alpha\beta}$ diverges on the CH. Thus it should come as no
surprise that
the Kretchsmann invariant, given by
\begin{eqnarray}
R_{\alpha\beta\gamma\delta}R^{\alpha\beta\gamma\delta} &\simeq&
8 r_{,VV} \!\!\left[ \frac{(r^2)_{,UU} - F_{,U} (r^2)_{,U}}{\alpha^2
e^{2F}r}\right]
 + \ldots\\
 &\sim& 16\frac{L_{\rm out}(U)L_{\rm in}(V)}{\alpha^2 e^{2F}
r^2}\label{eqn:22}
\end{eqnarray}
to leading order, also diverges like the measured energy density
provided that
the outflux from the star is non-vanishing.  This shows that a scalar
curvature
singularity  forms along the CH whenever a blueshifted influx and
continuous
outflux from the star are present.

\section{Conclusion}

For completeness, we will take the Ori model limit
of our
solution.  This is achieved by replacing
the
continuous outflux by a delta function source at $U=U_0$ and then
integrating
over $U$ where possible.  The result is
\begin{eqnarray}
r^2 &\simeq&  r_{\rm i}^2 -
2\int_{V_0}^{V}\!\!dV'\left[\int_{V_0}^{V'}\!\!dV''
L_{\rm out}(V'')\right]
+ \int_{U_0}^{U}\!\! \!e^{2F'}\left[ V - 2b
\right] dU'\: ,\label{eqn:ORI1}\\
m(V) &\simeq& {\rm const.} \times \int_{V_0}^{V}\!\!\left[ L_{\rm out}(V')
\right]dV'
- {\beta}
 \int_{V_0}^{V}\!\!\left[ V'L_{\rm out}(V') \right]dV'
 +\ldots \label{eqn:ORI2}
\end{eqnarray}
where $\left. b = e^{-F}\right|_{U=U_0}$, and
 is in complete agreement with~\cite{Ori:91}.  Since \mbox{$L_{\rm
out}(U)\equiv
0 $} for
$U>U_0$,
 it is clear that
the leading term in (\ref{eqn:22}) is absent,  and the divergence of the
curvature scalars is proportional
to the
mass function.  Therefore this model misses the leading behavior discussed
above,
but we wish to stress that the tidal forces~(and hence tidal distortion)
experienced by a free-falling observer in
both models are almost identical!

        In our opinion the result~(\ref{eqn:22}) is rather important.
It shows that
$L_{\rm out}(U)$ must be continuous in order to see the full nature of the
singularity
which forms along the CH.  Based on this observation it is tempting
to
speculate that asphericities may act to change the nature of the
singularity by
pushing divergences (proportional to $L_{\rm in}(V)$) into the Weyl
curvature.
Indeed,
asymptotic and perturbative analyses~\cite{perturb} of the
singularity inside a
more realistic black hole, suggest that the leading terms are in the
radiative
part of the Weyl curvature.  It seems to us very important, then, to
further
investigate the internal structure of black holes in aspherical
models.

 At the present time, the indications are that the singularity which forms
inside a realistic black hole is null.  A question which deserves some
attention is, how generic is this null picture?  Within spherical symmetry,
we
see that the singularity becomes spacelike when $r(U,V)\rightarrow 0$.  The
decrease in $r(U,V)$ (near the CH) is governed primarily by the outflux from
the star~[Eqn. (\ref{eqn:15})], and thus the null portion of the singularity
can be quite large.  It is possible (likely) that the presence of shear in
more
realistic models will change this, provoking a spacelike singularity which
is
asymptotically null near $P$ in Fig.~1.  This would be consistent with the
perturbative analysis in~\cite{perturb} and, as pointed out by
Yurtsever~\cite{Yurt}, would be similar to the situation in plane wave
spacetimes where perturbations do not capture the generic structure for
colliding waves.
 Finally we suggest that it is rather premature to discuss
  the strength of the singularity which
forms, or
the possible fate of an astronaut who falls into a black hole, since it is
likely that tidal
forces  will be substantially enhanced if the
singularity in more realistic models is spacelike.

Some of these issues are currently under investigation and details will be
presented elsewhere.

\acknowledgements
It is a great pleasure to thank Werner Israel for fruitful comments and
encouragement, and also Eric Poisson for constructive criticism which helped
us
clarify the exposition.  D. N. is grateful for the awards given by External
Affairs and International Trade Canada, Government of Canada Awards,
administrated by the International Council for Canadian Studies.  He also
thanks DGAPA,  UNAM for support.  This work was also partly supported by the
Natural Sciences and Engineering Research Council of Canada.
Figures~2, (i) and (ii), and some calculations in this work were done using
Mathematica$^{\copyright}$.

\appendix{The surface gravities}

In order to obtain the physical regions in the ab-plane in figure~2,
and to
examine the conditions on the surface gravities we found the
following useful.

Lake~\cite{Lake} introduced a rescaled quantity $x=r/m_1$, where
$m_1$ is the
mass of the black hole, so that the scalar function $f$ is given by
\begin{equation}
f=\frac{- b}{m_1^2 x^2}(x^4 - \frac{x^2}b + \frac{2 x}b -
\frac{a}b)\: .
\label{*}
\end{equation}
The surface gravity at any horizon $r$ is similarly given by
\begin{equation}
\kappa_r = \frac{1}{ m_1 b {x^3}_r}\left|{x^2}_r - 3 x_r + 2 a\right|
{}.
\label{**}
\end{equation}
By construction the roots of (\ref{*}) satisfy
$ 0 \leq x_{\rm i} \leq x_{\rm e} \leq x_{\rm c} $, and  we know that the
surface
gravities are
always greater than zero.  These facts and (\ref{**}) tell us that
\begin{equation}
 x_{\rm i} \leq \frac{3 - \sqrt{9 - 8 a}}2 \leq  x_{\rm e} \leq \frac{3 +
\sqrt{9 -
8 a}}2
\leq  x_{\rm c}\: .
\end{equation}
We now observe that for $a=9/8$, we have $ x_{\rm e}=3/2$ and since this
satisfies
(\ref{*}), $b=2/27$.  In fact for these values of a and b we have
$ x_{\rm i}= x_{\rm e}= x_{\rm c}$~\cite{Lake}.

Finally, the coalescence of the two or more roots occurs when
$\kappa_r =0$.
Using (\ref{*}) gives
\begin{equation}
b=\frac{y^2 -2y+a}{y^4} \: , \ \ \ \ y= \frac{3\pm\sqrt{9-8a}}{2}\: ,
\end{equation}
which allows us to plot the boundary of the regions of interest.

\figure{
The global structure of the spacetime, with crossflowing null dust.
EH is the
event horizon, CH is the Cauchy horizon and the cosmological horizon
is at
$r= r_{\rm c}$. The lines $U=U_0$ and $V=V_0$ at which the fluxes are
switched on are
also shown.}

\figure{The ab-plane.  Along curve 1 $ r_{\rm c}= r_{\rm e}$, and along
curve 2
$ r_{\rm i}= r_{\rm c}$.
(i)The region of physical interest is bounded by the axes and the
curves 1 and
2. (ii) A close up of the physical region. In the shaded portion no
mass
inflation occurs, but energy densities are divergent on the CH.
Above this,
both the mass function and the energy density diverge, and below is
the region
of complete stability }

\par\vfill\eject
\end{document}